\documentclass[10pt,conference]{IEEEtran}
\IEEEoverridecommandlockouts

\usepackage[utf8]{inputenc}
\usepackage[T1]{fontenc}     
\usepackage{cite}
\usepackage{amsmath,amssymb,amsfonts}
\usepackage{algorithm}
\usepackage{algpseudocode}
\usepackage{graphicx}
\usepackage{textcomp}
\usepackage[table]{xcolor}
\usepackage{booktabs}
\usepackage{multirow}
\usepackage{tikz}
\usetikzlibrary{arrows.meta,positioning}
\usepackage{url}
\usepackage{flushend}        
\usepackage[hidelinks]{hyperref}

\begin{document}

\title{Explainable Rule Mining of\\
       IPv6 Extension-Header Presence Patterns\\
       from Paired-Vantage Captures}

\author{%
\IEEEauthorblockN{%
Priyanka~Sinha\textsuperscript{1},
Nikolaos~Kekatos\textsuperscript{2},
Stylianos~Basagiannis\textsuperscript{3},
Antonio~Anastasio~Bruto~da~Costa\textsuperscript{1},\\
Alexios~Lekidis\textsuperscript{4},
Pabitra~Mitra\textsuperscript{1},
Tom~Nianios\textsuperscript{2},
Elpiniki~Papageorgiou\textsuperscript{4}%
}
\IEEEauthorblockA{%
\footnotesize
\parbox{0.95\textwidth}{\centering
\textsuperscript{1}IIT Kharagpur, India:\enspace
{\ttfamily\scriptsize \{priyanka.sinha.iitg,antonio.cse.iitkgp,pabitra\}@gmail.com}
\textbar{}
\textsuperscript{2}Clone Systems, CY:\enspace
{\ttfamily\scriptsize \{nkekatos,tnianios\}@clone-systems.com}
\textbar{}
\textsuperscript{3}IHU, GR:\enspace
{\ttfamily\scriptsize basagiannis@ihu.gr}
\textbar{}
\textsuperscript{4}Univ.\ of Thessaly, GR:\enspace
{\ttfamily\scriptsize \{alekidis,elpinikipapageorgiou\}@uth.gr}%
}}}

\maketitle
{\renewcommand{\thefootnote}{}\footnotetext{\copyright~2026 IEEE. Personal use of this material is permitted. Permission from IEEE must be obtained for all other uses, in any current or future media, including reprinting/republishing this material for advertising or promotional purposes, creating new collective works, for resale or redistribution to servers or lists, or reuse of any copyrighted component of this work in other works. This work has been accepted for publication at the 2026 IEEE International Conference on Cyber Security and Resilience (IEEE CSR 2026).}}

\begin{abstract}
IPv6 extension headers (EHs), such as fragmentation, segment
routing, and in-situ telemetry, are operationally important yet
widely dropped in transit, and characterising their behaviour from
packet captures is a recurring measurement problem. We ask whether
an explainable miner can recover human-readable rules of EH
behaviour, and we contribute two reusable tools: a
\emph{negative-control protocol} that diagnoses whether a mined
``temporal'' network rule reflects genuine cross-packet dynamics or
mere within-packet co-occurrence, and a sender-conditioned,
per-family EH-retention measurement. Applying an interpretable
temporal-logic rule miner to the JAMES paired-vantage dataset, we
recover a portable Fragment-EH rule that the protocol reveals to be
a within-packet, near-definitional co-occurrence rather than a
temporal pattern, so the temporal-logic machinery does no work for
this dominant rule; the retention measurement independently recovers
the expected within-window ordering of EH observability. Our
main result is therefore an honest, controlled negative finding,
corroborated by executed decision-tree and large-language-model
baselines: on the evaluated JAMES traces \emph{network-temporal}
structure does not carry the dominant Fragment-EH signal, and we
supply the controls that establish when it would, validated on a
synthetic positive control containing a genuine cross-packet dependency.
\end{abstract}


\IEEEpeerreviewmaketitle

\section{Introduction}

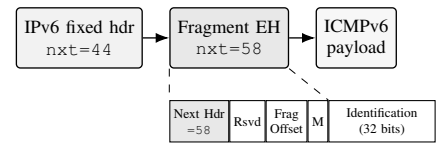
\begin{figure}[b]
\centering
\begin{tikzpicture}[
  every node/.style={font=\scriptsize},
  hdr/.style={draw, rounded corners=1.5pt, line width=0.5pt,
    minimum height=8mm, align=center, inner sep=3pt},
  fld/.style={draw, line width=0.4pt, minimum height=5.5mm,
    align=center, inner sep=1.5pt, font=\tiny},
  arr/.style={-{Latex[length=1.8mm]}, line width=0.5pt}]
  \node[hdr, fill=black!5] (a) {IPv6 fixed hdr\\\texttt{nxt=44}};
  \node[hdr, fill=black!9, right=3.5mm of a] (b) {Fragment EH\\\texttt{nxt=58}};
  \node[hdr, fill=black!3, right=3.5mm of b] (c) {ICMPv6\\payload};
  \draw[arr] (a.east) -- (b.west);
  \draw[arr] (b.east) -- (c.west);
  \node[fld, fill=black!9, below=7mm of b.south west, anchor=west] (f1) {Next Hdr\\\texttt{=58}};
  \node[fld, right=0pt of f1] (f2) {Rsvd};
  \node[fld, right=0pt of f2] (f3) {Frag\\Offset};
  \node[fld, right=0pt of f3] (f4) {M};
  \node[fld, right=0pt of f4, text width=13mm] (f5) {Identification\\(32 bits)};
  \draw[dashed, line width=0.3pt] (b.south west) -- (f1.north west);
  \draw[dashed, line width=0.3pt] (b.south east) -- (f4.north east);
\end{tikzpicture}
\caption{An IPv6 header chain (real capture). The fixed header's
Next-Header ($\mathtt{nxt}{=}44$) points to a Fragment EH, whose
Next-Header ($58$) points to the ICMPv6 payload (box: its $8$ bytes);
a middlebox must walk the chain and many instead drop the packet. Our
predicates read these markers off the wire: \texttt{HAS\_FRAG\_EH}
flags a Fragment header in the chain, and as JAMES probes carry a
single EH the fixed-header Next-Header identifies it directly.}
\label{fig:ehchain}
\end{figure}

IPv6~\cite{rfc8200} is the current version of the Internet
Protocol, the scheme that addresses and routes every packet on the
Internet. It was created chiefly because its predecessor, IPv4, was
running out of addresses; besides a vastly larger address space,
IPv6 was given a simpler and \emph{extensible} header. Instead of
packing optional features into a fixed header, IPv6 carries them in
a chain of \emph{extension headers} (EHs) placed between the main
$40$-byte header and the data (Fig.~\ref{fig:ehchain}). Each header
points to the next through a Next-Header field, so anything
inspecting the packet must read down the chain. Different EHs do
different jobs: fragmenting oversized packets, source routing and
Segment Routing (SRv6), carrying diagnostics such as Performance and
Diagnostic Metrics (PDM)~\cite{rfc8250} and telemetry, and IPsec
security. This design lets operators add network functions without
changing the base protocol.

The catch is that many devices along the path, such as firewalls
and routers, do not fully process this variable-length chain and, a
great deal of the time, simply \emph{drop} packets that carry
extension headers. Because Internet standards are
voluntary~\cite{rfc8890}, real deployments diverge
from the written specification~\cite{rfc8953}, and EHs are among the
starkest examples: since RFC~7872, repeated Internet-wide
measurements have found intermediate nodes silently discarding
EH-bearing packets, at rates that vary by network operator
(Autonomous System, AS), by EH type, and over
time~\cite{rfc7872,hendriks-eh,james-imc2022}. Anything that relies
on EHs, from diagnostics to tunnelling, therefore cannot be trusted
to work end-to-end without first measuring
it. We treat this as a
measurement-and-method problem and are careful \emph{not} to claim a
security consequence our data do not show: EH drop is fail-closed,
and we do not assert an exploitable attack surface here.

For a network operator this is a practical problem: before turning
on an EH-dependent service, one needs to know \emph{which} extension
headers survive \emph{on which paths}, and today that answer comes
from bespoke, throw-away measurement scripts. What would help is a
way to take a batch of captured packets and automatically produce a
short, \emph{human-readable} description of which packet features go
with EH presence or absence, and how that varies across measurement
points, so the same result both predicts and explains. This is the
goal of \emph{explainable} machine
learning~\cite{rudin-stop-explaining}: instead of a black box,
produce a model whose output \emph{is} the explanation. Equivalently
it is a \emph{specification-mining} problem~\cite{acm-formal-mining},
learning the rule from observed data rather than writing it by hand.
A readable rule is only useful, though, if it reflects real
behaviour and not an artefact of how the packets happen to be built,
which is exactly what we guard against.

PSIMiner~\cite{psiminer} fits this formulation. It learns rules in
a fragment of metric temporal logic (MTL)~\cite{alur-henzinger} as
\emph{timed sequences of predicates}, with inter-event intervals
that it can, in principle, refine from data. It has previously been
used to explain outcomes of multi-party
dialogues~\cite{sinha2021dialogues}; we apply the same pipeline to
network packet traces. We deliberately restrict the predicate
vocabulary to IPv6-fixed-header fields and EH-chain flags: this
keeps every mined rule expressible in terms a network engineer
reads directly off the IPv6 header, which is the explainability
property the design rests on.

\textbf{Contributions.}
(i)~We cast IPv6 extension-header \emph{presence} analysis as timed
temporal-logic specification mining over labelled \texttt{pcap}
captures, with an IPv6-header-only vocabulary as a deliberate
explainability constraint (\S\ref{sec:methodology}).
(ii)~Our reusable methodological contribution is a
\emph{negative-control protocol} (joint-row permutation,
independent-column shuffle, lag-augmented decision-tree) that
diagnoses whether a mined ``temporal'' rule is genuine cross-packet
dynamics or within-packet co-occurrence (\S\ref{sec:robustness}).
(iii)~The headline is a controlled \emph{negative finding}: across
all $21$ JAMES paired-vantage receivers the single portable
Fragment-EH rule
(\texttt{!PLEN\_MED |=> \#\#[0:4] !HAS\_FRAG\_EH}, $87.4$--$98.9\%$)
is a within-packet, near-definitional co-occurrence, so the
temporal-logic machinery does no work for this dominant rule.
(iv)~Two executed baselines, a decision tree and an LLM rule-miner,
reproduce this (\S\ref{sec:llm-baseline}).
(v)~A sender-conditioned retention measurement recovers an ordering
consistent with prior JAMES findings (\S\ref{sec:survivability}).

\section{Methodology}
\label{sec:methodology}

PSIMiner~\cite{psiminer} learns properties in a fragment of
real-time temporal logic~\cite{alur-henzinger} restricted to the
future operator $\Diamond_{[a,b]}$, conjunctions of predicates over
packet attributes, and Boolean implication:
\begin{equation}
\varphi := \phi \;\wedge\; \Diamond_{[a,b]} \varphi'
\,\mid\, \phi, \qquad \phi : S \to S,
\label{eq:mtl}
\end{equation}
where $S$ is a Boolean expression over a finite set of ground
predicates on the trace attributes. This fragment captures
properties of the form ``after event $A$, within $[a,b]$ time units
event $B$ occurs,'' including chained sequences with
non-deterministic intervals, but stops short of full MTL: the
restriction keeps mining tractable while preserving an interpretable
surface syntax.

Given a labelled trace corpus and a target outcome predicate $E$,
specification mining learns candidate $\varphi$ such that
$\varphi \rightarrow E$ holds frequently enough in the data to be
useful. The learned $\varphi$ is itself the explanation: unlike
post-hoc attribution, the mined rule is a global, human-readable
property of the data, not a local saliency map over a black box.

\subsection{Pipeline overview}
The pipeline has three stages:
(1)~parse each \texttt{pcap} into a comma-separated-value (CSV)
time series of attributed packet events; (2)~choose a target
predicate that varies inside each trace, such as an IPv6 EH presence
flag like \texttt{HAS\_FRAG\_EH} ($\mathtt{nxt}=44$), together with a
small set of ground predicates over packet attributes; (3)~run
PSIMiner on the corpus and inspect the mined rules. The mined rule
is the explanation; there is no post-hoc explainer and no language
model in the mining core. Algorithm~\ref{alg:pipeline} summarises
the procedure.

\begin{algorithm}[t]
\caption{EH-presence rule mining pipeline}
\label{alg:pipeline}
\begin{algorithmic}[1]
\Require Capture set $\mathcal{P}$; target predicate $E$;
PSIMiner template parameters $(n,k,d,\ell)$
\Ensure Ranked timed rules
$\{\varphi_i \Rightarrow E\}$ with correlations $\alpha_i$
\For{$p \in \mathcal{P}$}
    \State $T_p \gets \Call{TsharkProject}{p}$
           \Comment{per-packet CSV}
    \State $\hat{T}_p \gets \Call{Bucketize}{T_p}$
           \Comment{Boolean predicates}
\EndFor
\State $\mathcal{R} \gets
       \Call{PSIMiner}{\{\hat{T}_p\}_{p},\, E;\; n, k, d, \ell}$
\State \Return $\Call{SortByCorrelation}{\mathcal{R}}$
\end{algorithmic}
\end{algorithm}

\subsection{Trace representation}
We parse each capture with \texttt{tshark} and project it onto a
fixed per-packet schema with the relative timestamp in the first
column. To keep the analysis strictly within the IPv6 header
(the explainability constraint), we use only IPv6-fixed-header
fields and IPv6 Next-Header chain markers; upper-layer protocol
facts and measurement-process facts are intentionally excluded.
\begin{itemize}
\item \texttt{nxt}: IPv6 Next Header value flagging presence of
each EH (categorical; $0$ Hop-by-Hop, $43$ Routing, $44$ Fragment,
$50$ ESP, $51$ AH, $60$ Destination Options).
\item \texttt{plen}: IPv6 Payload Length (bytes), bucketed.
\item \texttt{hlim}: IPv6 Hop Limit, bucketed.
\item \texttt{flow}: IPv6 Flow Label, as a zero/non-zero binary.
\end{itemize}
Each EH type becomes its own binary predicate
(\texttt{HAS\_HBH\_EH}, \texttt{HAS\_RT\_EH}, \texttt{HAS\_FRAG\_EH},
\texttt{HAS\_ESP\_EH}, \texttt{HAS\_AH\_EH}, \texttt{HAS\_DST\_EH});
presence is never bunched into a single \texttt{HAS\_EH} flag.
Payload Length is bucketed using protocol-semantic thresholds: a
small ICMPv6 echo without an EH has a $56$-byte payload; with a
fragment header it sits in larger buckets, and the SRv6 header
expands the IPv6 portion by $24+$ bytes. We expose four
\texttt{plen} predicates (\texttt{PLEN\_SMALL} $\leq\!64$\,B,
\texttt{PLEN\_MED} $\geq\!128$\,B, \texttt{PLEN\_BIG}
$\geq\!512$\,B, \texttt{PLEN\_HUGE} $\geq\!1280$\,B), two
\texttt{hlim} predicates (\texttt{HLIM\_LOW} $\leq\!16$,
\texttt{HLIM\_HIGH} $\geq\!200$), and one \texttt{flow} predicate
(\texttt{FLOW\_ZERO}). The cut-points are protocol-driven rather
than data-driven quantiles; \S\ref{sec:threats} discusses the
trade-off. Note that the vocabulary contains \emph{no} AS or
vantage predicate: the pipeline cannot, by construction, emit an
AS- or provider-conditioned rule, and we do not claim it can
(\S\ref{sec:motivation}).

\subsection{Temporal pattern mining with PSIMiner}
\label{sec:methodology-causation}
PSIMiner is parameterised by an \emph{explanation template}
\begin{equation}
B_n \,\#\#[0{:}k]\, \cdots \,\#\#[0{:}k]\, B_0
   \;\longmapsto\; E,
\label{eq:template}
\end{equation}
where each $B_i$ is a \emph{bucket} of zero or more predicates, $n$
and $k$ are meta-parameters, and $E$ is the target. The template
has $n+1$ buckets separated by intervals of up to $k$ time units.
Read in English: ``if $B_n$ holds, and within $0$ to $k$ time units
$B_{n-1}$ holds, and so on until $B_0$, then $E$ holds.'' During
learning PSIMiner decides which predicates populate which buckets
and may leave buckets empty. Each output rule is annotated with a
\emph{correlation} metric~\cite{psiminer}. It measures, for
$S \Rightarrow E$, a \emph{target-coverage}
(recall) quantity, the fraction of $E$-observations that have a
matching $S$ in the antecedent window, i.e.\ approximately
$P(S \mid E)$, \emph{not} the implication confidence $P(E \mid S)$.
The two differ sharply on this corpus: every Fragment-EH packet is
medium-payload, so $P(\mathtt{PLEN\_MED}\mid\mathtt{HAS\_FRAG\_EH})
=1.00$ (coverage) while only
$P(\mathtt{HAS\_FRAG\_EH}\mid\mathtt{PLEN\_MED})=0.81$ (confidence).
We therefore report coverage and confidence separately
(\S\ref{sec:llm-baseline}) and never read a high coverage as an
implication. Correlation is a data-fit measure over the received
trace; \S\ref{sec:robustness} returns to what it does and does not
certify. The recurring $93.04$/$87.1$/$97.2\%$ are one within-packet
link scored differently: native coverage of the \emph{negative} rule
\texttt{!MED$\Rightarrow$!FRAG}, the \emph{positive} forward rule
\texttt{MED$\Rightarrow$FRAG}, and a validator re-score, respectively.

\paragraph{Correlation vs.\ causation}
We use ``causal'' only in PSIMiner's sense: temporal precedence plus
statistical association (Granger~\cite{granger}), not intervention or
counterfactuals (Pearl~\cite{pearl-causality}); the mined rules are
correlational statements conditional on temporal ordering, and any
causal reading rests on the measurement campaign's design, not the
mining algorithm.

\subsection{Public datasets}
\label{sec:datasets}
We rely exclusively on publicly available IPv6 EH data.

\paragraph{amwalding IPv6 EH pcap collection~\cite{amwalding-eh}}
A curated public set of \texttt{pcapng} fixtures covering
Hop-by-Hop, Fragmentation, Routing/SRv6, and ESP. Volumes are small
($1$ to $65$ packets/file) but the byte-level layouts are real; we
use it as a sanity substrate (\S\ref{sec:results-pub}).

\paragraph{JAMES measurement dataset~\cite{james-imc2022}}
An Apache-2.0, paired-vantage IPv6 EH corpus (ULi\`ege; IMC 2022)
across $21$ controlled Internet-wide vantage cities, whose
dual-vantage structure makes EH stripping directly observable.
JAMES ships per-vantage \texttt{received\_traffic.pcap} captures
($\approx$$35$--$40$k packets/receiver) and per-source experiment
timestamps documenting which EH variants were sent
(\texttt{frag512m0}, \texttt{frag1280m0}, \texttt{RoutingType0}--$6$,
\texttt{Dest*skip}, \texttt{HbH*skip}, etc.). Our pipeline ingests
the receiver pcaps directly via the \texttt{tshark} projection.

\section{Results}
\label{sec:results}

We report: (i)~the pipeline runs end-to-end on the amwalding
substrate (\S\ref{sec:results-pub}); (ii)~at measurement scale on
the JAMES receiver pcaps it produces a single portable
high-correlation rule that the controls in \S\ref{sec:robustness}
identify as \emph{within-packet} (\S\ref{sec:results-james}).
We report mined rules verbatim as
$\textsc{antecedent} \mathrel{|\!\!=\!\!>} \#\#[a:b]\;
\textsc{consequent}$; \emph{correlation} is PSIMiner's data-fit
metric.

\subsection{Pipeline on the amwalding substrate}
\label{sec:results-pub}
The amwalding fixtures ($1$--$65$ packets/file, $79$ combined) run
end-to-end with the IPv6-only vocabulary, parsing every file and
ingesting all EH types present (Hop-by-Hop, Routing, Fragment, ESP).
With so few packets per file the mined rules are unstable and
setup-dependent, so we treat this substrate purely as a \emph{sanity
check} that the pipeline parses public EH data correctly and draw no
measurement-scale conclusion; all quantitative results below are on
the JAMES corpus.

\subsection{Mining at scale on JAMES receiver pcaps}
\label{sec:results-james}
We run the same pipeline on the JAMES
\texttt{received\_traffic.pcap} captures. After \texttt{tshark}
projection AMS and TYO become time series of $38{,}692$ and
$35{,}031$ packets carrying a mix of EHs in operationally relevant
volumes (at AMS: $11{,}838$ Destination-Options, $10{,}369$
Routing, $7{,}783$ Fragment packets). We use
$\mathtt{seqLength}=4$, $\mathtt{depth}=5$,
$\mathtt{bestPredCount}=8$, $\mathtt{delayRes}=1.0$.

\paragraph{A single antecedent, portable across all 21 vantages}
With target \texttt{HAS\_FRAG\_EH} the dominant rule is the same on
every JAMES receiver:
\begin{quote}\small\ttfamily
!PLEN\_MED |=> \#\#[0:4] !HAS\_FRAG\_EH
\end{quote}
i.e.\ packets whose payload is not medium-or-larger are followed
within $4$ seconds by no Fragment EH. The pipeline mines the identical antecedent across all \emph{$21$} JAMES receiver
vantages (four continents, $757{,}932$ packets), with correlation in the band
$87.4$--$98.9\%$. \S\ref{sec:robustness} shows \emph{why} the rule
is so portable: it is a within-packet regularity that is
near-definitional on this corpus, so it recurs wherever Fragment
probes are received.

\paragraph{``By construction'': the antecedent is near-definitional}
On the AMS receiver, \emph{all} $7{,}783$ Fragment-EH packets
($\mathtt{nxt}=44$) carry $\mathtt{plen}\geq128$ (minimum observed
payload $458$\,B); $P(\mathtt{HAS\_FRAG\_EH})=0.201$. This is a
\emph{sender-side} property of how JAMES forges probes: Fragment
probes are generated at $\geq512$\,B
(\texttt{frag512m0}/\texttt{frag1280m0}), not a property of the
network path. Consequently \texttt{!PLEN\_MED $\Rightarrow$
!HAS\_FRAG\_EH} is a near-definitional \emph{within-packet}
statement: because the antecedent is satisfied on the same packet as
the (negated) consequent, the $\#\#[0{:}4]$ interval is not doing
any work.

\paragraph{Auxiliary targets}
Mining \texttt{HAS\_HBH\_EH} returns no rule above threshold;
Hop-by-Hop is absent at every vantage except Sydney ($509$ packets,
${<}1.5\%$), consistent with the JAMES finding that it is the most
aggressively dropped category. Mining
\texttt{HAS\_DST\_EH}/\texttt{HAS\_RT\_EH} surfaces negative rules
whose antecedents are \emph{other} EH flags, reflecting the JAMES
schedule of testing one EH variant per window, i.e.\ the temporal
dimension of the data is the \emph{probe clock}, not network
dynamics.

\section{Robustness, Baselines, and Retention}
\label{sec:robustness}
We assess the JAMES-scale rule along several axes that together form
the paper's reusable methodological contribution: a protocol that
tells whether a mined ``temporal'' rule is genuine cross-packet
dynamics or within-packet structure.
Table~\ref{tab:taxonomy} organises the analysis into five rule sets
under one reporting scheme, progressing from same-packet structure
to paired-vantage retention; we then detail each control below.

\begin{table}[t]
\centering
\caption{Consolidated rule taxonomy (AMS receiver). \emph{Score} is
confidence (RS1 same-packet, RS5 window-conditioned) or \emph{forward ordered
correlation} (RS2--RS4), the fraction of consequent events whose
antecedent occurs in the stated preceding interval under the original
packet order; \emph{perm} is the joint-row-permuted score ($20$ draws)
and $\Delta$ is score minus perm.}
\label{tab:taxonomy}
\footnotesize
\begin{tabular*}{\columnwidth}{@{}l@{\extracolsep{\fill}}rrr@{}}
\toprule
\textbf{Rule} & \textbf{score} & \textbf{perm} & \textbf{$\Delta$} \\
\midrule
\rowcolor{gray!15}\multicolumn{4}{@{}l}{\textit{RS1 same-packet structure (confidence): within-packet}}\\
FRAG\,$\to$\,MED & 100.0 & --- & --- \\
MED\,$\to$\,FRAG & 80.6 & --- & --- \\
\rowcolor{gray!15}\multicolumn{4}{@{}l}{\textit{RS2 MED $\Rightarrow$ FRAG delay sweep (fwd): within-packet\,+\,probe-burst}}\\
\quad \#\#[0:4] (incl.\ same-pkt) & 87.1 & --- & --- \\
\quad \#\#[$\epsilon$:4] & 86.9 & --- & --- \\
\quad [1pk:4pk] (cross-pkt) & 79.2 & 59.3 & $+20$ \\
\rowcolor{gray!15}\multicolumn{4}{@{}l}{\textit{RS3 persistence (fwd): measurement-temporal (bursty)}}\\
\quad FRAG\,$\Rightarrow$\,[1pk:4pk]\,FRAG & 96.6 & 59.3 & $+37$ \\
\rowcolor{gray!15}\multicolumn{4}{@{}l}{\textit{RS4 cross-EH succession (fwd): measurement-temporal (probe clock)}}\\
\quad DST $\Rightarrow$ RT & 12.2 & 91.7 & $-80$ \\
\quad RT $\Rightarrow$ FRAG & 10.8 & 83.4 & $-73$ \\
\rowcolor{gray!15}\multicolumn{4}{@{}l}{\textit{RS5 sender-window retention (conf): paired-vantage}}\\
\quad DST & 55.3 & --- & --- \\
\quad RT & 47.4 & --- & --- \\
\quad FRAG & 44.3 & --- & --- \\
\quad HBH & 0.0 & --- & --- \\
\bottomrule
\end{tabular*}

\smallskip
{\scriptsize Scores in \% (RS1/RS5 confidence; RS2--RS4 forward
correlation). MED\,=\,PLEN\_MED; EH names abbreviate HAS\_$\ast$\_EH;
pk\,=\,packet lag. Order-independent (within-packet) rules have
$\Delta\!\approx\!0$; large $|\Delta|$ marks order-dependence, which
here traces to the JAMES probe schedule. \emph{perm} is shown only
for the cross-packet RS2 [1pk:4pk] (the others mix same-packet and
window effects); it falls to the base rate ($59.3\%$): probe-burst,
not within-packet.}
\end{table}

Table~\ref{tab:taxonomy} shows this directly: the headline
size--fragment rule is order-independent (within-packet,
$\Delta\!\approx\!0$), while RS3, RS4, and even the strictly-cross-packet
RS2 variant are order-dependent and trace to the probe schedule, not
network dynamics. We detail the controls below.

\paragraph{Joint-row permutation control}
For each receiver we randomly permute the complete per-packet
IPv6-header tuple
$(\mathtt{nxt},\mathtt{plen},\mathtt{hlim},\mathtt{flow})$ across
rows (a \emph{joint-row} permutation: it preserves the within-packet
joint distribution and marginal frequencies but destroys temporal
ordering), then re-mine. The top rule \emph{does not} collapse: at
AMS the re-mined correlation rises from $93.04\%$ to $100.00\%$ and
at TYO from $94.31\%$ to $96.46\%$. To confirm this is not a
single-draw artefact, we re-evaluate the fixed rule's validator
correlation over $20$ independent permutations: it is
$100.0\pm0.00\%$ for the headline within-packet rule (invariant),
against an ordered value of $97.2\%$, whereas a cross-EH rule such
as \texttt{HAS\_FRAG\_EH |=> !HAS\_DST\_EH} moves from $42.3\%$
(ordered) to $99.6\pm0.08\%$ (permuted). This rise is expected: the
JAMES schedule blocks EH families in time, so cross-family
successions are rare in order but permutation creates artificial
adjacency that inflates coverage (the mechanism behind RS4's negative
$\Delta$). A rule that is \emph{invariant} to
temporal order is a within-packet co-occurrence, not a temporal
pattern.

\paragraph{Independent-column shuffle}
A stronger control permutes each IPv6-header predicate column
\emph{independently}, destroying the within-packet
$\mathtt{HAS\_FRAG\_EH}\leftrightarrow\mathtt{PLEN\_MED}$
association. Over $20$ independent such shuffles the
size$\to$fragment confidence collapses from $0.806$ to
$0.200\pm0.003$ (the base rate), and PSIMiner returns no rule above
threshold. The two controls localise the explanatory signal
precisely: it lives in the within-packet IPv6 geometry, and
\emph{only} there.

\paragraph{Rule complexity vs.\ correlation: does temporal structure
pay off?}
PSIMiner's template natively expresses multi-bucket
(cross-packet) sequences, so we searched for the best rule at each
antecedent complexity, i.e.\ number of temporal buckets, across all
six EH targets. The strongest rule by
far is single-bucket and within-packet ($93.04\%$). The best
\emph{genuinely} two-bucket rule,
\texttt{!HAS\_DST\_EH \#\#[0:1] PLEN\_MED |=> !HAS\_RT\_EH}, reaches
only $24.8\%$, and the best three-bucket rule only $2.3\%$; no
four-bucket rule scores above the two-bucket value (a released
script regenerates the search, each target completing in
$3$--$27$\,s after retrying the non-deterministic crash). Correlation
collapses as the rule becomes genuinely temporal: the more a rule
relies on cross-packet timing, the weaker it is on this corpus.

The order-dependent rules are real, so the machinery is not inert
\emph{in general}: under permutation the RS3/RS4 rules
(Table~\ref{tab:taxonomy}) swing by $37$--$80$\,pp, against
$\Delta\!\approx\!0$ for the within-packet headline. As a
\emph{positive control}, a synthetic trace in which a marker
deterministically causes a target two packets later ($38{,}692$ rows,
marker prevalence $0.20$, fixed seed, $20$ permutations) scores $99.99\%$
ordered but collapses to the base rate ($59.0\%$, $\Delta\!=\!+41$)
under permutation, so the protocol correctly \emph{flags} a genuine
cross-packet rule, not only rejects one. Permutation alone gives the
order-independent/order-dependent split; telling measurement-temporal
(the probe clock) from network-temporal rests on the probe schedule.
Here every order-dependent effect is measurement-temporal; the
network-temporal regime is absent.

\paragraph{Decision-tree cross-check}
As a finite-window baseline we train a scikit-learn decision tree
(DT) on the AMS receiver with the same bucketised vocabulary. A
depth-$5$ per-row tree reaches $96.93\%$ full-data accuracy; the
number generalises rather than being a training-only
artefact: at a chronological $70/30$ split (train first, test last,
in time order) test accuracy is $98.1\%$ (precision $0.93$, recall
$1.00$, positive class \texttt{HAS\_FRAG\_EH}), and a stratified
$5$-fold cross-validation gives $96.9\pm0.2\%$. Extending the
features with row-lagged predicates at $K\in\{1,2,4\}$ (the
closest analogue to a finite context window) makes \emph{no}
marginal difference (chronological test accuracy $0.981\to0.978$ as
$K$ grows): the tree finds no useful split on a lagged feature. The
within-packet geometry already saturates the task, so temporal
context is inert: not evidence our method beats a tree, but
\emph{confirmation of the negative result}.

\paragraph{One-dimensional hyperparameter ablation}
We sweep each of $(\mathtt{seqLength},\mathtt{depth},
\mathtt{delayRes})$ around the baseline $(4,5,1.0)$ on AMS with the
other two fixed (Table~\ref{tab:ablation}). The antecedent
$\neg\mathtt{PLEN\_MED}$ is invariant at every grid point.
Correlation is flat in $\mathtt{depth}$ ($93.04\%$ throughout) and
rises gently with $\mathtt{seqLength}$ and $\mathtt{delayRes}$. The
mined interval upper bound is \emph{exactly}
$\mathtt{seqLength}\times\mathtt{delayRes}$ at every cell
(Table~\ref{tab:ablation}). The learned interval is thus
\emph{hyperparameter-determined}, not data-refined, on this
corpus: PSIMiner's data-refinement is a \emph{capability} the
within-packet corpus does not exercise, so nothing about the interval
is learned from cross-packet timing here.

\paragraph{On the PSIMiner segfaults}
The public PSIMiner~$1.1$ build exhibits a \emph{non-deterministic}
crash (SIGSEGV): the same configuration completes on some runs and
crashes on others (even the baseline crashed once), so it is a tool
memory-safety bug independent of the grid point, not a signal about
any configuration. We retry each cell until it completes; every cell
in Table~\ref{tab:ablation} is populated within $1$--$3$ attempts
and the grid has no holes. A hardened build is engineering future
work.

\begin{table}[t]
\centering
\caption{Hole-free one-dimensional ablation on the AMS receiver,
$\mathtt{HAS\_FRAG\_EH}$ target, IPv6-only vocabulary. Baseline
$(\mathtt{seqLength},\mathtt{depth},\mathtt{delayRes})=(4,5,1.0)$ in
bold. The mined interval upper bound equals
$\mathtt{seqLength}\times\mathtt{delayRes}$ exactly
(hyperparameter-determined). All cells completed on retry (the
PSIMiner~$1.1$ crash is non-deterministic).}
\label{tab:ablation}
\small
\begin{tabular}{@{}llrrl@{}}
\toprule
\textbf{Axis} & \textbf{Value} & \textbf{corr (\%)} & \textbf{supp (\%)} & \textbf{mined $\#\#[0{:}k]$} \\
\midrule
\multirow{4}{*}{$\mathtt{seqLength}$}
  & 2 & 92.28 & 79.78 & $\#\#[0{:}2]$ \\
  & 3 & 92.66 & 80.11 & $\#\#[0{:}3]$ \\
  & \textbf{4} & \textbf{93.04} & \textbf{80.44} & $\#\#[0{:}4]$ \\
  & 5 & 93.40 & 80.75 & $\#\#[0{:}5]$ \\
\midrule
\multirow{4}{*}{$\mathtt{depth}$}
  & 3 & 93.04 & 80.44 & $\#\#[0{:}4]$ \\
  & 4 & 93.04 & 80.44 & $\#\#[0{:}4]$ \\
  & \textbf{5} & \textbf{93.04} & \textbf{80.44} & $\#\#[0{:}4]$ \\
  & 6 & 93.04 & 80.44 & $\#\#[0{:}4]$ \\
\midrule
\multirow{3}{*}{$\mathtt{delayRes}$}
  & 0.5 & 92.28 & 79.78 & $\#\#[0{:}2]$ \\
  & \textbf{1.0} & \textbf{93.04} & \textbf{80.44} & $\#\#[0{:}4]$ \\
  & 2.0 & 94.41 & 81.62 & $\#\#[0{:}8]$ \\
\bottomrule
\end{tabular}
\end{table}

\paragraph{Cross-vantage bootstrap}
Resampling the $n=21$ per-vantage correlations with replacement ($B=10{,}000$) gives a
bootstrap $95\%$ interval on the mean of $[93.60\%, 96.08\%]$ (mean
$94.86\%$, s.d.\ $2.89$): the within-packet rule has a central
tendency in the mid-nineties across the JAMES set rather than being
driven by one capture. This interval describes variability
\emph{within} the $21$ JAMES receivers; it is not a population-level
confidence interval over Internet vantage points, ASes, or
deployments.

\subsection{Executed baseline: LLM rule-miner}
\label{sec:llm-baseline}
The decision tree isolates the \emph{propositional} half of a
generative-plus-logical miner such as NetNomos~\cite{netnomos}. As
its \emph{generative} complement we run an LLM rule-miner and score
its proposals with an \emph{independent} validator. This baseline is
\emph{executed}, not illustrative: candidates come from a live model
and every correlation is computed deterministically by
\texttt{rule\_validator.py} on the full trace, so the numeric column
is reproducible from the released artefacts (only the model's
free-text proposals carry run-to-run variance).

\paragraph{Setup}
We prompt Claude Opus~$4.8$ with (i) the $13$-predicate IPv6-header
vocabulary, (ii) the rule template, and (iii) a $200$-row uniform
random sample of the AMS trace (fixed seeds $2026$--$2028$). The
model returns $1$--$3$ candidate rules in a strict JSON schema; we
validate each on the full $38{,}692$-row trace.

\paragraph{Findings}
The model
consistently proposes the \emph{polar inverse} of the PSIMiner rule,
\texttt{PLEN\_MED |=> \#\#[0:k] HAS\_FRAG\_EH}, at $100\%$ coverage
(again the within-packet $\mathtt{plen}\!\ge\!128$ link, not an
implication: its confidence is only $0.81$). An \texttt{HLIM\_HIGH}
rule validates at $97.57\%$ and a conjunctive rule at $95.41\%$. The
model also proposes \texttt{HAS\_DST\_EH |=> \#\#[0:4] HAS\_FRAG\_EH},
which the validator \emph{rejects} at $20.75\%$ (Destination-Options
and Fragment probes run in disjoint windows), exactly the
plausible-but-wrong guess the verifier exists to filter. The LLM
converges on the same within-packet size$\leftrightarrow$fragment
link, reinforcing the negative finding.

These baselines are \emph{complementary diagnostics}, not a
head-to-head benchmark (accuracy vs.\ correlation/support are not a
shared metric); their shared message is that three independent
methods converge on the same within-packet explanation.

\subsection{Sender-window-conditioned retention}
\label{sec:survivability}
Toward \emph{survival given a known sender configuration}, we joined
the JAMES per-source probe labels to the AMS receiver trace (each CSV
row is $1{:}1$ with a pcap frame; absolute time
$t+\min(\text{epoch})$ reconstructed to within $1.6\!\times\!10^{-7}$\,s of the original frame time). This
is \emph{not} a full survival analysis: packets are not matched
individually, and only $14{,}163/38{,}692=36.6\%$ fall inside a probe
window (reliable only at the $\approx$$90$\,s family-block
granularity). The window-conditioned retention
$P(\text{recv}=x\mid\text{sent}=x)$ (RS5, Table~\ref{tab:taxonomy})
produces an ordering consistent with prior JAMES observations:
Destination Options $55.3\%$, Routing $47.4\%$, Fragment $44.3\%$,
Hop-by-Hop $0\%$, each with lift ${>}1.7$; the apparent
\texttt{SENT\_DST}$\to$\texttt{RECV\_FRAG} ``transformation'' has lift
$0.42$, i.e.\ none (IPv6 fragmentation is source-side, not in
transit). Adding \texttt{SENT\_*} predicates still leaves the
within-packet rule \texttt{!PLEN\_MED |=> \#\#[0:4] !HAS\_FRAG\_EH}
($93.0\%$) dominant, sender labels entering only low-correlation
($4.1\%$), negligible-support rules: conditioning on what was sent
adds no predictive lift, so even this yields a within-packet,
presence-level result. True per-probe survival, needing packet-level
matching, remains future work.

\section{Discussion}
\label{sec:discussion}

\subsection{Threats to validity}
\label{sec:threats}
The amwalding fixtures are tiny; the rule recovered there is a
sanity check, not a measurement finding. Predicate cut-points are
protocol-driven, not data-driven quantiles, which keeps them stable
but ties them to the EH geometry. Most importantly, the central rule
is a \emph{within-packet, partly sender-imposed} regularity
(\S\ref{sec:results-james}, \S\ref{sec:robustness}); its
$87$--$99\%$ portability reflects that near-definitional link, not a
discovered temporal pattern, and the ``temporal'' axis of this
corpus is the JAMES probe clock. The evidence is also from a single
controlled-probe corpus (JAMES); our MAWI check
(\S\ref{sec:motivation}) found EHs essentially absent from natural
traffic, so external validity, and generality to EH targets beyond
Fragment, is future work. As in
\S\ref{sec:methodology-causation}, the mined rules report
correlation under temporal precedence, not interventional causation.

\subsection{Related work and future directions}
\label{sec:motivation}
The provider-level observations that motivate this work, e.g.\ the
per-AS Routing-Header drop tables of the JAMES IETF
draft~\cite{vyncke-v6ops-james} (DigitalOcean/OVH on the RH0/RH4
lists, Vultr clean, hyperscalers uncovered), are the draft's
findings, not products of our AS-agnostic pipeline; we use them as
motivation, not results. Our work spans five threads: EH
survivability measurement (RFC~7872;
JAMES~\cite{james-imc2022,iurman-razor}), which asks \emph{whether}
EHs survive where we ask whether their \emph{presence} carries
structure; logic/invariant mining from traces
(NetNomos~\cite{netnomos}); temporal-logic mining and runtime
verification~\cite{psiminer,kekatos-rv,frehse-kekatos};
interpretable-by-design models~\cite{rudin-stop-explaining}; and
association-rule metrics~\cite{agrawal-assoc}, whose conflation of
co-occurrence with implication is exactly why we separate coverage,
confidence, and lift and use permutation controls. NetNomos differs
from us in \emph{representation} (data-refined intervals vs.\ a
fixed window), which does not matter here since the signal is within-packet.

\paragraph{Future work}
The concrete next step is a full sender-labelled survival campaign
restricted to probe windows (\S\ref{sec:survivability}). For
external validity we re-ran the pipeline on \emph{natural} backbone
traffic. In the MAWI archive~\cite{mawi} (samplepoint-F, file
\texttt{202512311400.pcap.gz}, a $15$-min backbone capture read with
\texttt{tshark -Y ipv6}), the $2025$-$12$-$31$ trace held $8.2$M IPv6 frames but only $490$ Hop-by-Hop and $18$ Fragment
EHs and \emph{no} Routing, Destination-Options, ESP, or AH headers:
extension headers are vanishingly rare (about $0.006\%$) in this
sampled backbone trace, so EH behaviour must be measured by
\emph{active} probing, as JAMES does. A hyperscaler (Google/Azure/AWS) campaign
instantiating the cloud EH-testing methodology~\cite{elkins-cloud}
would close the coverage gap; an $8$-capture Vultr-vs-Google PDM
pilot is only a preliminary observation. An LLM-driven rendering
pass could phrase mined rules against the relevant RFCs
(RFC~8200~\cite{rfc8200}, RFC~8250~\cite{rfc8250}).

\section{Conclusion}
\label{sec:conclusion}
We studied an explainable pipeline that mines IPv6 extension-header
presence patterns from packet traces as short timed temporal-logic
rules, where the mined rule is the explanation. On all $21$ JAMES
paired-vantage receivers the pipeline recovers one portable
antecedent for Fragment-EH presence ($87.4$--$98.9\%$), but our
negative-control protocol shows this is a \emph{within-packet}
co-occurrence that is near-definitional by JAMES probe
construction, so the metric-temporal-logic machinery is inert for the
dominant Fragment-EH rule on this corpus. A decision tree and an executed LLM rule-miner
independently reproduce the finding. The contribution we stand
behind is therefore twofold: a controlled negative result
(network-temporal structure does not carry the dominant Fragment-EH signal on these
real traces), and the reusable negative-control protocol
(joint-row permutation, independent-column shuffle, lag-augmented
tree) that
diagnoses exactly when a mined ``temporal'' network rule is genuine
and when it is within-packet structure in disguise.

\section*{Acknowledgment}
This work has received funding from the European Union's Digital
Europe Programme under grant agreement No~101190251 (CYBERGUARD).
Code and the derived per-vantage CSVs are available at
\url{https://github.com/nikos-kekatos/ipv6-eh-rule-mining}.

\end{document}